\documentclass{rspublic}

\newcommand\ba{\begin{eqnarray}}
\newcommand\ea{\end{eqnarray}}
\newcommand\bc{\begin{center}}
\newcommand\ec{\end{center}}
\newcommand\ra{\rightarrow}
\def\bi{\begin{itemize}}
\def\ei{\end{itemize}}
\def\bn{\begin{enumerate}}
\def\en{\end{enumerate}}
\def\bmp{\begin{minipage}}
\def\emp{\end{minipage}}

\newcommand\p{\partial}

\def\epsilon{\varepsilon}
\def\e{\mathrm{e}}
\def\sech{\ensuremath{\mathrm{sech}}}
\def\e{\ensuremath{\mathrm{e}}}
\def\i{\ensuremath{\mathrm{i}}}

\usepackage[dvips]{graphicx}
\usepackage{amsmath}
\usepackage{amsfonts}
\usepackage{latexsym}

\begin{document}
\title[Coherent structures for localized states]{The emergence of a
coherent structure for coherent structures:
localized states in nonlinear systems}

\author{J. H. P. Dawes}

\affiliation{Department of Mathematical Sciences, University of Bath,
Claverton Down, Bath BA2 7AY, UK}

\label{firstpage}
\maketitle

\begin{abstract}{pattern formation, Turing instability,
bifurcation, homoclinic snaking}
Coherent structures emerge from the dynamics of many kinds of
dissipative, externally driven, nonlinear systems, and continue
to provoke new questions that challenge our
physical and mathematical understanding.

In one specific sub-class of such problems, where
a pattern-forming, or `Turing', instability occurs,
rapid progress has been made recently in our
understanding of the formation of localized states: patches of
regular pattern surrounded by the unpatterned homogeneous
background state.

This short review article surveys the progress that has been made for
localized states, proposes three areas of application for
these ideas that
would take the theory in new directions and ultimately be of substantial
benefit to areas of applied science.
Finally I offer speculations for future work, based on localized states,
that may help researchers to understand coherent structures more generally.

\end{abstract}


\section{Introduction}
\label{sec:intro}

Encouraged by the success of low-dimensional dynamical
systems theory in the 1980s in explaining the origin of complicated
behaviour in nonlinear ordinary differential equations,
researchers in what might be called `nonlinear science' are
attempting to carry through a similar programme for spatially-extended
systems of many kinds. In this review I will restrict attention
to systems that are internally dissipative and externally driven;
`nonequilibrium' in the physics terminology. More precisely, such
systems (for example
chemical and biological kinetics, viscous fluid mechanics, frictional
solid mechanics) equilibrate at a level of activity sufficient to
provide a global energy balance. Commonly this involves the formation
of spatial structure with a single well-defined lengthscale. The
bifurcations in which such 
a patterned state is born out of a homogeneous state are known,
naturally enough, as spontaneous pattern-forming,
or `Turing' instabilities (Turing 1952). Spontaneity
is important here: the external driving is still imposed
in a spatially uniform fashion, so the emergence of structure
is a clear symmetry-breaking transition.

Well-known examples of spontaneous pattern formation
include Rayleigh--B\'enard convection,
in a layer of viscous fluid heated from below,
Faraday surface waves
on a the surface of a vertically shaken liquid or granular layer,
and reaction-diffusion dynamics, proposed for example to organise many
processes in developmental biology (Cross \& Hohenberg 1993;
Murray 2003; Hoyle 2006).

The theoretical analysis of `pattern-forming instabilities' of these
kinds in fluid mechanical problems can be traced back
to the early 20th century in the case of
Rayleigh-B\'enard convection (Rayleigh 1916).
The subsequent emergence of mathematical biology as a distinct
area within applied mathematics has meant that Alan Turing's demonstration
(Turing 1952) that such patterns could result from a
more general, and biologically-relevant, mechanism of local activation
coupled with longer-range inhibition 
has continued to have a significant
impact in shaping our understanding of mechanisms for morphogenesis.

Turing's original observation, that spontaneous pattern formation
may arise through a linear instability due to
coexisting diffusive effects that operate on sufficiently different
spatial scales even when the homogeneous state is stable in the absence
of diffusion, has been made mathematically precise, and forms the
basis for arguments about the generic nature of such
instabilities and the typical patterns which result.

More recently, it has become clear that a pattern-forming or
`Turing' instability can also result in the formation of
spatially localized states, even if the system remains driven in a 
spatially uniform fashion. These localized states resemble a number
of periods of the periodic pattern that we might expect, but
surrounded by the spatially uniform background state rather than extending
to fill the whole spatial domain. The two key ingredients for the formation
of localized states near Turing instabilities are
{\it (i) bistability}, and
{\it (ii) pinning}.

{\it Bistability}
means that the spatially uniform state and the patterned state are
stable to
small disturbances over a single range of values of the system parameters.
In the language of bifurcation theory, bistability occurs when the
Turing instability is subcritical, creating a small-amplitude
unstable pattern which exists alongside a larger amplitude
stable pattern as well as the stable trivial unpatterned state.

{\it Pinning} refers, by analogy with the motion of defects in crystals,
to the local energy well in which the localized state sits: there
is an energetic barrier to overcome in order either
for the localized patch
of pattern to propagate further into the surrounding background state, or
conversely, for the background state to be able to swamp the
localized state and remove it. In spatially continuous systems, described
by partial differential equations, the periodicity of the pattern
itself provides this pinning effect: pinning is therefore a generic feature
for these systems. In spatially discrete dissipative systems,
where the discrete
nature of the system is modelled in ways that are closely
analogous to those used in atomic lattices (such as the Frenkel--Kontorova
model), the discrete
nature of the system generates pinning also in a generic fashion.

Recent mathematical work has proved that localized states arise generically
near Turing instabilities, and has greatly
clarified their existence and bifurcations in
model equations such as the 1D bistable Swift--Hohenberg equation:
\ba
u_t & = & r u - \left( 1 + \p_x^2\right)^2 u + N(u;s),
\label{eqn:sh}
\ea
(Swift \& Hohenberg 1977)
where $u(x,t)$ is a scalar variable and $r<0<s$ are parameters, and
$N(u;s)$ refers to the choice of nonlinear terms that give rise
to the subcritical bifurcation of small amplitude states
and subsequence re-stabilisation
of the dynamics at larger amplitudes. Popular choices are
$N_1(u;s)=su^2-u^3$ and $N_2(u;s)=su^3-u^5$. The resulting bifurcation
structure of localized states has a characteristic structure containing
two intertwining curves of solutions and as a result the process of
formation of these families of localized states is frequently referred
to as `homoclinic snaking'.
While it is not a normal form in the strict sense,~(\ref{eqn:sh})
is often taken in the literature as a canonical model
equation for homoclinic snaking, and recent investigation through a
combination
of numerical and analytic approaches supports its employment as 
a generic model equation.

The structure of this review is as follows.
In section~\ref{sec:recent} I list
physical systems in which localized states have been observed, either
in laboratory or numerical experiments, beforel discussing
analytical approaches
to the dynamics of~(\ref{eqn:sh}). Section~\ref{sec:apps} presents
three areas in which both bistability and pinning effects
are likely to be present and hence the theory of
localized states should be an important part of our understanding
of the problem. Conclusions and wilder speculations
for future research directions are contained in section~\ref{sec:conc}.

\section{Recent progress}
\label{sec:recent}

In this section I first summarise a number of areas in which
localized states have been observed, either in laboratory
experiments, or in theoretical modelling work
(section~\ref{sec:recent}\ref{sec:motiv}). Then 
I very briefly outline the dynamics of the canonical
1D Swift--Hohenberg model~(section~\ref{sec:recent}\ref{sec:sh}),
before discussing additional issues.

\subsection{Motivations}
\label{sec:motiv}

Localized patterns have been described in
experiments and models in an extremely wide variety of fields.
The literature on the localisation of buckling patterns
of elastic beams and shells is particularly rich
(Potier--Ferry 1983, Hunt et al. 2000); localized
states have also been analysed in
fluid mechanics (in particular in doubly-diffusive
convection problems (Riecke \& Granzow 1999; Riecke 1999;
Batiste \& Knobloch 2005) and
magnetoconvection (Blanchflower 1999; Dawes 2007, 2008);
nonlinear optics (Akhmediev \& Ankiewicz 2005);
gas discharge systems (Purwins et al. 2005);
vertically oscillated granular and viscoelastic media
(Umbanhowar et al. 1996, Lioubashevski et al. 1999);
ferrofluid instability (Richter \& Barashenkov 2005),
surface catalysis, mathematical
neuroscience, developmental biology and many others.

The study of localized states is therefore of fundamental
importance to research in all these fields.
However, it is only recently that the mathematical structure and
organisation of localized states has become completely
understood even in the simplest, one-dimensional, case.
Alongside this mathematical structure, recent work
has exploited the existence of numerical continuation packages
such as AUTO (Doedel 2007). The resulting bifurcation diagrams that
can be compiled
clarify hugely the organisation, existence, and stability of localized
patterns and have contributed to a resurgence of interest in
the area.

\subsection{The Swift--Hohenberg model}
\label{sec:sh}

The simplest pattern-formation situation 
in which localized states appear
is given by considering a single scalar PDE for a quantity $u(x,t)$
that is posed on the real line $-\infty < x <
\infty$. We assume that the trivial state $u(x,t) \equiv 0$ exists for
all parameter values and is linearly stable when a parameter $r$ 
is negative. We further assume that it first loses stability, 
at $r=0$, to Fourier modes $\e^{\i k x}$ with $k$ near unity.
We take the PDE to be first-order in time
and left-right reflection symmetric, i.e. unchanged under the
operation $(x,u) \ra (-x,u)$. This latter condition
(a `reversibility') implies that
terms in the PDE contain even numbers of $x$-derivatives. The
simplest canonical model equation with these properties
is the cubic--quintic fourth-order Swift--Hohenberg equation
\ba
u_t & = & ru - (1+\partial_x^2)^2 u + s u^3 - u^5. \label{eqn:sh35}
\ea
For $s>0$ the instability of the state $u(x,t)=0$ at $r=0$ is to modes
$\sim \e^{\i k x}$ with $k$ near unity, and this instability
is subcritical. As a result, there is a region of bistability between
stable large-amplitude space-periodic solutions and the trivial solution $u=0$,
see figure~\ref{fig:snake1}(a).

Equation~(\ref{eqn:sh35}) can be analysed from
two points of view. One is the asymptotic reduction
of~(\ref{eqn:sh35}) to a Ginzburg--Landau equation via a multiple-scales
expansion. The second is to consider equilibria of
the fourth-order spatial dynamical system in $x$
given by setting $u_t=0$ in which localized states correspond to
trajectories homoclinic to $u=0$.

\subsubsection{Asymptotics}

\begin{figure}[!h]
\bc
\includegraphics[width=6.25cm]{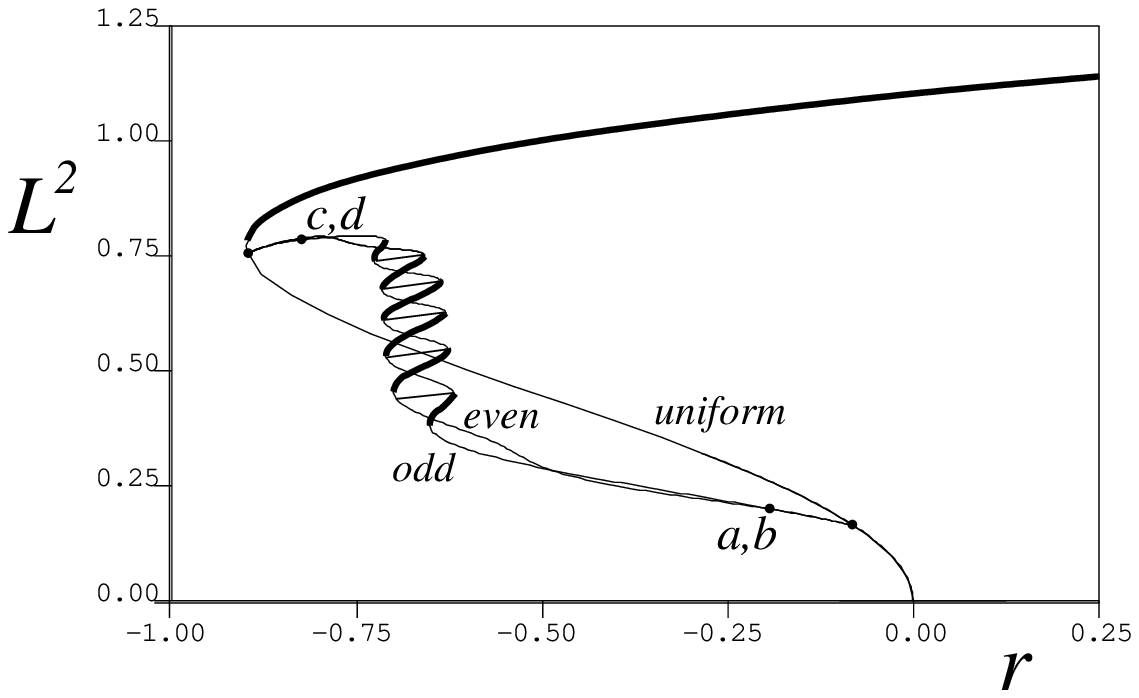}
\includegraphics[width=6.25cm]{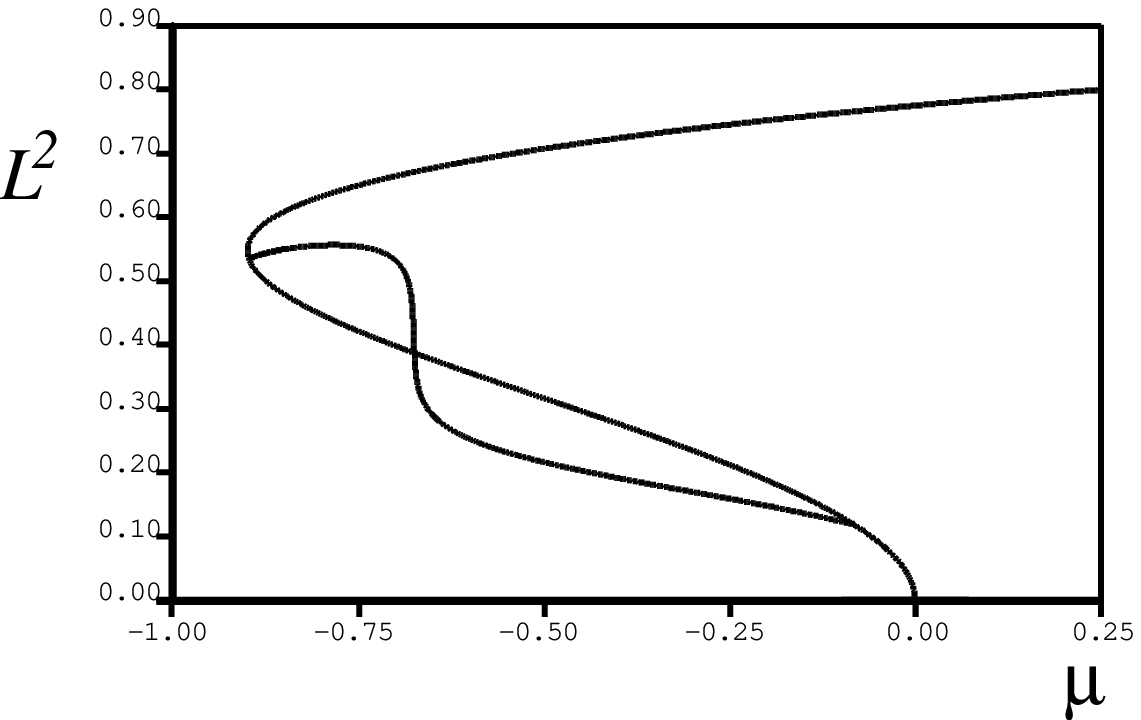}
\centerline{(a) \hspace{6.0cm} (b)}
\caption{\label{fig:snake1}Bifurcation diagrams for
(a) the Swift--Hohenberg equation~(\ref{eqn:sh35}) solved in the domain
$0\leq x \leq L=10\pi$
and (b) the Ginzburg--Landau
equation~(\ref{eqn:gl}) solved in the domain $0\leq X \leq 10\pi$, both
using periodic boundary conditions. The vertical line
segment of the curve in (b) indicates the Maxwell point $\mu_{mx}$
around which the snaking curves in (a) open up and intertwine.}
\ec
\end{figure}

In the Ginzburg--Landau approach near the codimension-two
point where $r=s=0$ it is appropriate to introduce the
scaled variables $\mu$ and $\hat{s}$ defined by
$r=\epsilon^2 \mu$, $s=\epsilon^2 \hat{s}$, and
the long length and time scales $X=\epsilon^2 x$ and $T=\epsilon^4 t$.
We look for steady solutions to~(\ref{eqn:sh35}) perturbatively with the
multiple-scales expansion
$u(x,t) = \epsilon \left( A(X,T)\e^{\i x} + c.c.
\right) + \epsilon^2 u_2(x,X,t,T) + O(\epsilon^3)$ where $c.c.$
denotes the complex conjugate.
We substitute this \textit{ansatz} into~(\ref{eqn:sh35}) and 
solve at successive orders in $\epsilon$, applying a `solvability
condition' at each order to eliminate
secular terms that would otherwise disrupt the asymptotic ordering of the
solution in powers of $\epsilon$. From the solvability
condition at $O(\epsilon^5)$ we deduce the cubic--quintic
Ginzburg--Landau equation for the complex-valued envelope $A(X,T)$:
\ba
A_T & = & \mu A + 4 A_{XX} + 3\hat{s} A|A|^2 - 10 A|A|^4. \label{eqn:gl}
\ea
Two distinct constant solutions $A=A_0^\pm$ exist for
$-9\hat{s}^2/40<\mu<0$ with the smaller (larger) amplitude state being
unstable (stable), respectively. These correspond to exactly periodic
solutions for $u(x)$, see figure~\ref{fig:snake1}(b).
\begin{figure}[!h]
\bc
\includegraphics[width=3.0cm]{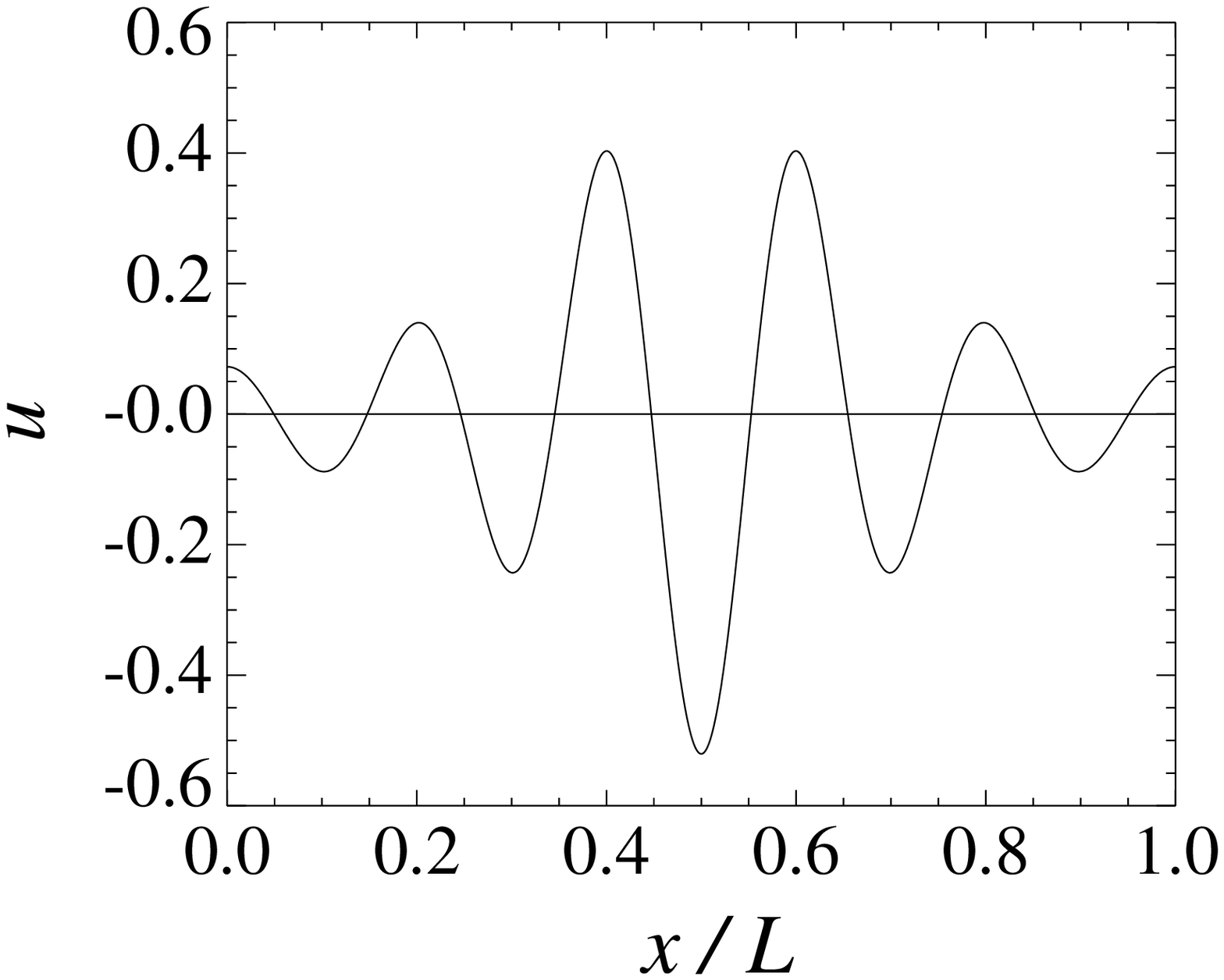}
\includegraphics[width=3.0cm]{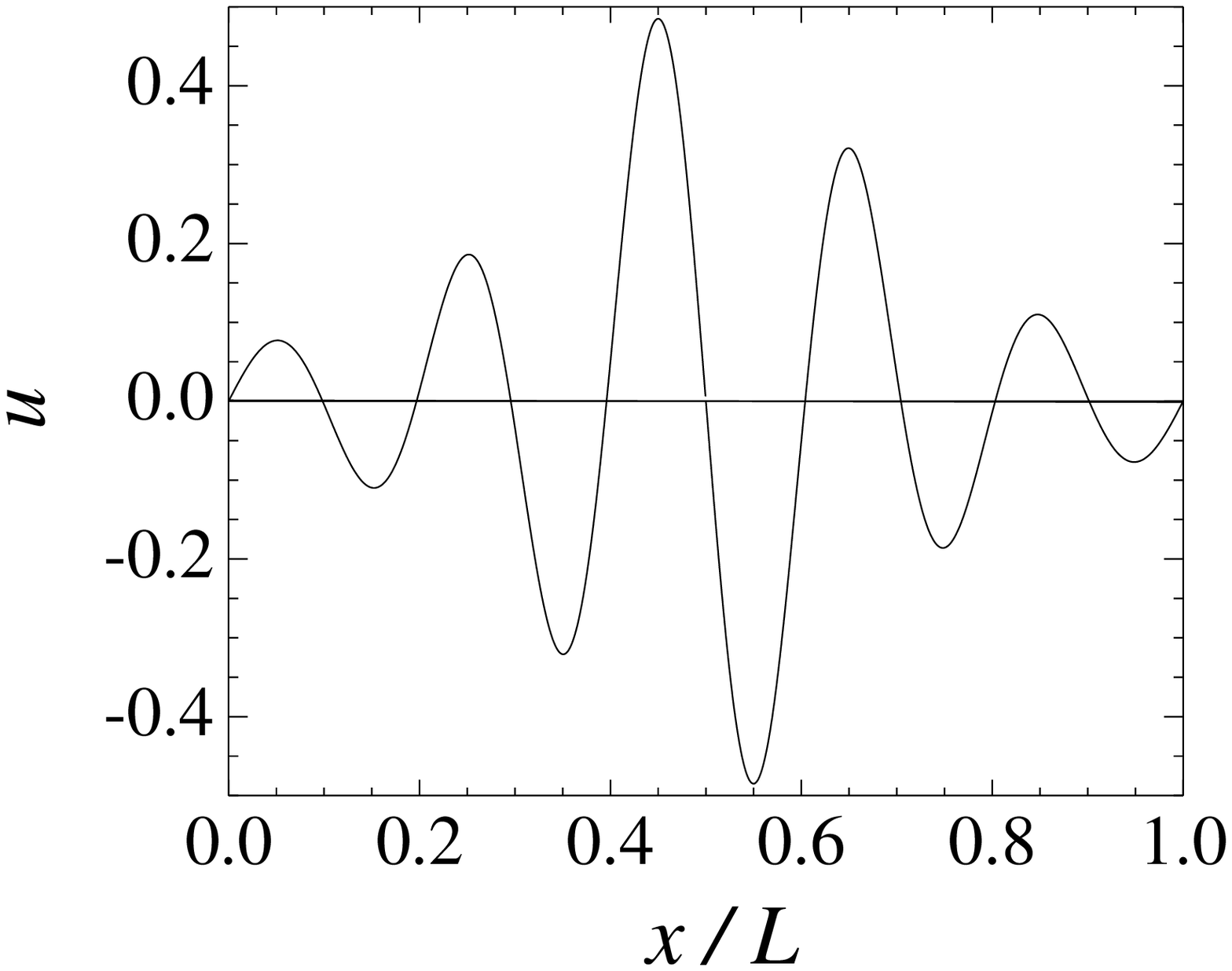}
\includegraphics[width=3.0cm]{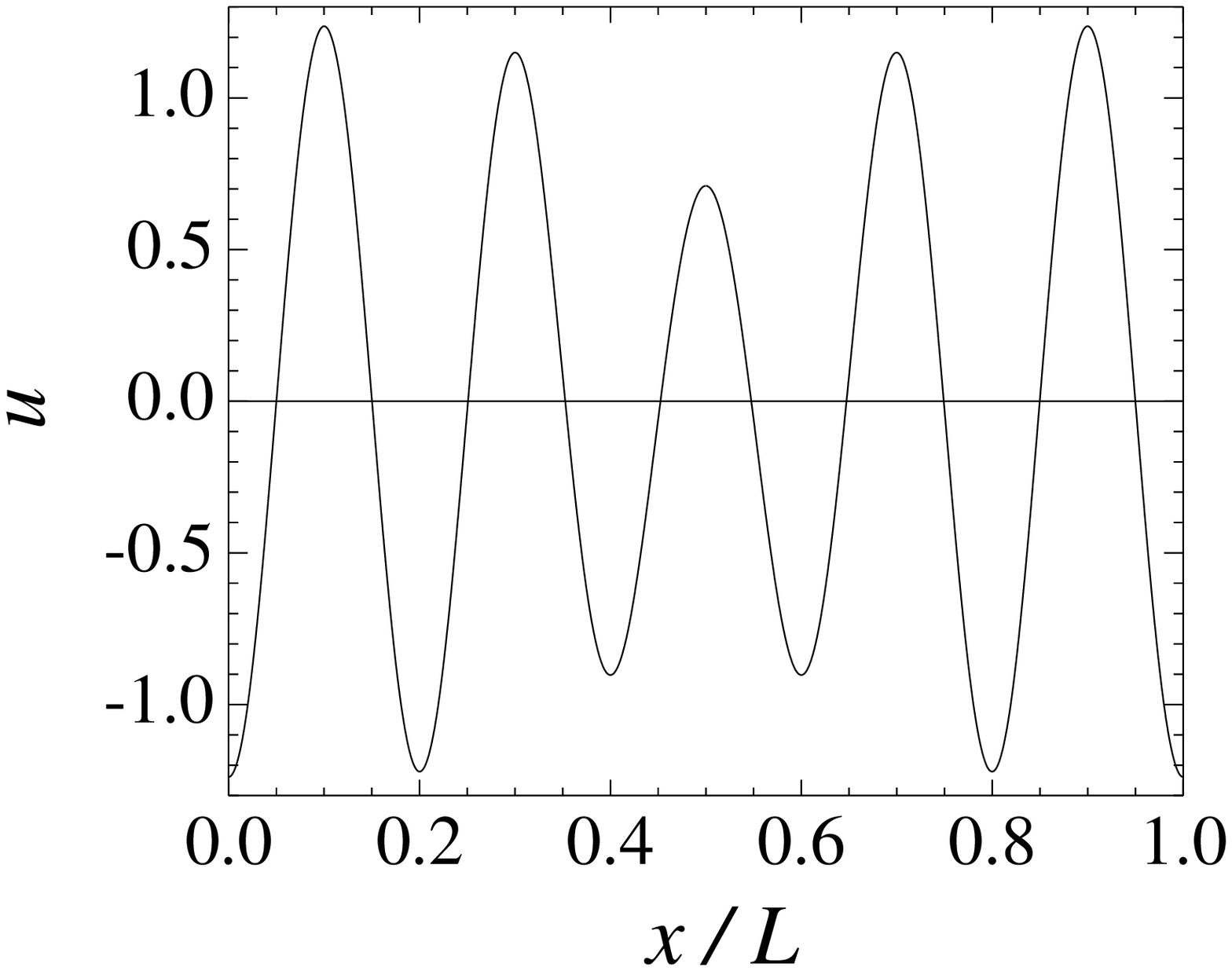}
\includegraphics[width=3.0cm]{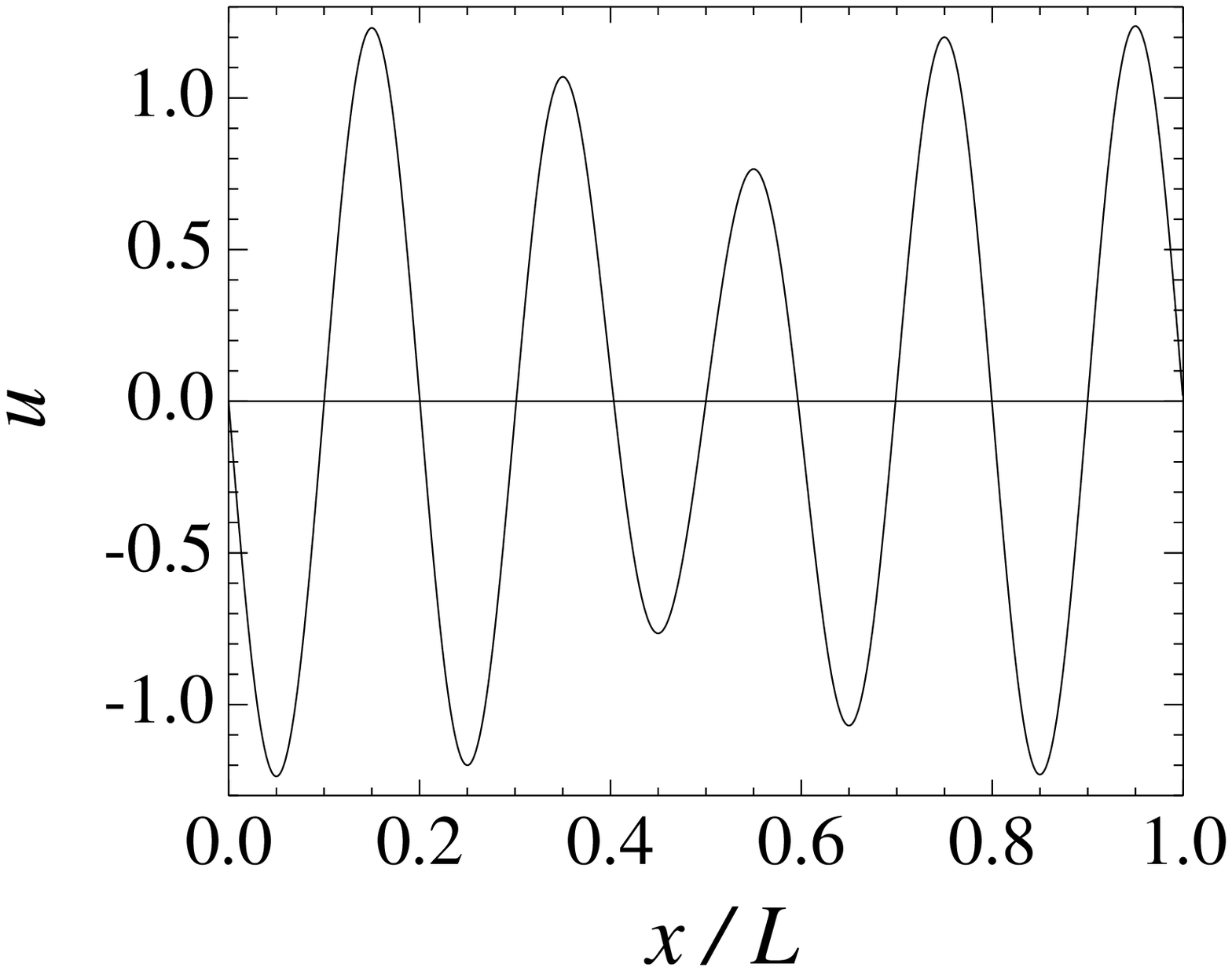}
\centerline{\phantom{aaa} (a) \hspace{2.4cm} (b) \hspace{2.4cm} (c) \hspace{2.4cm} (d)}
\caption{\label{fig:snake2}Localized states on the odd and even snaking curves
shown in figure~\ref{fig:snake1}(a), corresponding to the labels $a$, $b$,
$c$ and $d$. $(a)$ and $(c)$ lie on the even-symmetric
branch and $(b)$ and $(d)$ on the odd-symmetric branch.}
\ec
\end{figure}

There are also non-constant solutions
which correspond to localized states:
At small amplitudes $|A| \ll 1$ the balance between the first three
terms on the right-hand side of~(\ref{eqn:gl})
indicates that non-constant
equilibrium solutions also exist, with a profile that is close
to a `$\sech$' function. As $\mu$ decreases, the solution profile broadens
and more closely resembles a pair of `$\tanh$'-like fronts connecting
the constant solutions $A=A_0^+$ and $A=0$.
At the \textit{Maxwell point} $\mu=-27\hat{s}^2/160$ these two
constant solutions are energetically equal (note
that equations~(\ref{eqn:gl}) and~(\ref{eqn:sh35}) are
variational) and so a front between them remains stationary. Hence
it looks possible that stationary solutions for $u(x)$ can be
constructed which consist of a patch of almost uniform pattern
surrounded by the trivial state $u=0$.
That this indeed occurs rests on Pomeau's observation (Pomeau 1986)
that such
a front between the periodic pattern and the trivial solution $u=0$
is \textit{pinned} by the periodicity of the pattern itself. The
pinning effect emerges through the interaction of the long-wavelength
envelope scale $X$
and the original pattern scale $x$. Since these are decoupled
at every order in the multiple-scales perturbation theory, this
interaction must necessarily be a `beyond-all-orders' effect.
The relevant exponentially small terms,
discussed qualitatively by many authors,
have only recently been calculated correctly (Chapman \& Kozyreff 2009).
Pinning expands the region of existence of the
localized states from a single line into a
cusp-shaped wedge in the $(r,s)$ plane.
Within the cusp-shaped pinning region
the two curves of localized states intertwine in a characteristic
fashion which gives rise to the term `homoclinic snaking',
as shown in figure~\ref{fig:snake1}(a). For the cubic--quintic
Swift--Hohenberg equation~(\ref{eqn:sh35}) localized
states on the two curves are odd and even respectively,
as illustrated in figure~\ref{fig:snake2}.
The short horizontal lines in figure~\ref{fig:snake1}(a)
between the snaking curves represent `ladder' branches
of asymmetric (i.e. neither odd nor even) localized states.

\subsubsection{Spatial dynamics}

A different approach to~(\ref{eqn:sh}) is to neglect the
time derivative and take the spatial coordinate to be the time-like
evolution variable; this framework is therefore
often referred to as 
`spatial dynamics'. In this framework the pattern-forming instability
that occurs at $r=0$ is, due to the reversibility,
a Hamiltonian--Hopf bifurcation
(a.k.a. a $1:1$ resonance).
The normal form analysis of this Hamiltonian--Hopf bifurcation was
carried out by Iooss \& Perou\`eme (1993)
and later extended by Woods \& Champneys (1999), Coullet et al
(2000) and Burke
\& Knobloch (2006); taken together, these papers establish many features
of the bifurcation problem, for example the existence in the normal form
of small-amplitude localized states near the bifurcation
point and that two of these states are then guaranteed to exist along
bifurcating solution branches away from $r=0$: it is these branches that
correspond to the commonly identified localized patterns. 
The characteristic intertwined wiggling of the snaking branches
arises from the generic behaviour of the unstable and stable
manifolds of the origin as they pass through a heteroclinic tangle.

In the resulting bifurcation diagram
the secondary branches are sometimes referred to as
`ladders'. That such asymmetric branches should exist generically
can be deduced both from a general bifurcation-theoretic
approach (Beck et al. 2009)
and deduced from the exponential asymptotics results of
Chapman \& Kozyreff (2009).

\subsection{Effects of a finite domain}

The `spatial dynamics' approach to homoclinic
snaking necessarily deals with 
homoclinic orbits: solutions on the real line. Practical applications,
though, necessarily
demand finite domains and suitable boundary conditions. Numerical
investigation of the Swift--Hohenberg equation in
finite domains (Bergeon et al. 2008, Dawes 2009)
illustrates how the homoclinic snaking
persists when the domain is large and finite, and how the snake
winds up or unravels as the domain 
size increases or decreases, respectively.

As illustrated by figure~\ref{fig:snake1}(a), in a finite domain the
snaking curves bifurcate directly from the periodic pattern close
to, but not exactly at, $r=0$. As the localized state expands to fill
the domain it ceases to be localised and eventually
the snaking curve reconnects to the periodic pattern near
the saddle-node point on the periodic branch.
The details of this reconnection are
more complicated than one might perhaps expect, particularly
when a second parameter, for example the domain size $L$, is allowed
to vary. It appears that the reconnection of the snaking branch
is organised by a mode interaction between the $n$-roll and $n+1$-roll
branches of periodic patterns which move close to each other as
$L$ increases from $2\pi n$ to $2\pi (n+1)$ (Dawes, 2009).

\begin{figure}[!h]
\bc
\includegraphics[width=6.25cm]{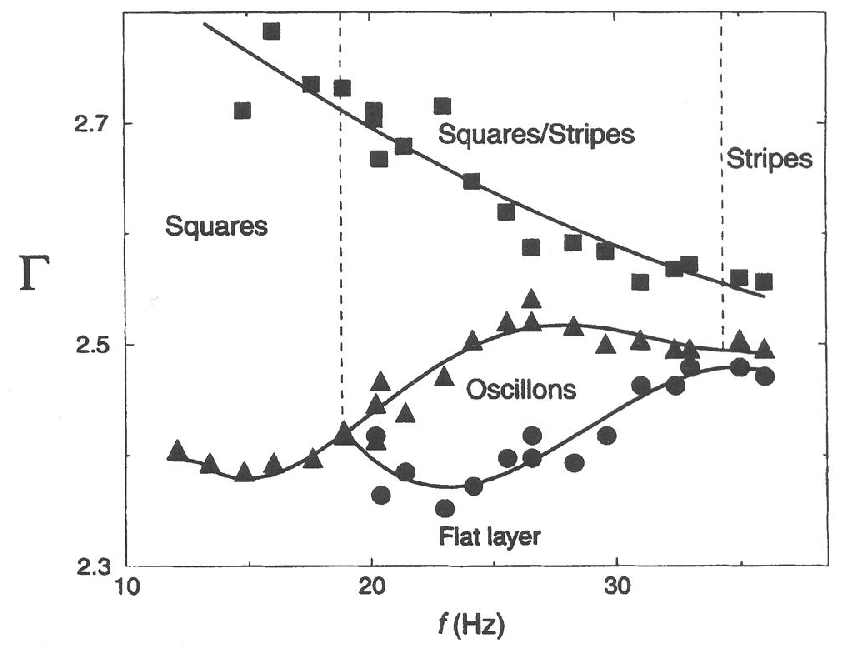}
\includegraphics[width=6.25cm,height=5.0cm]{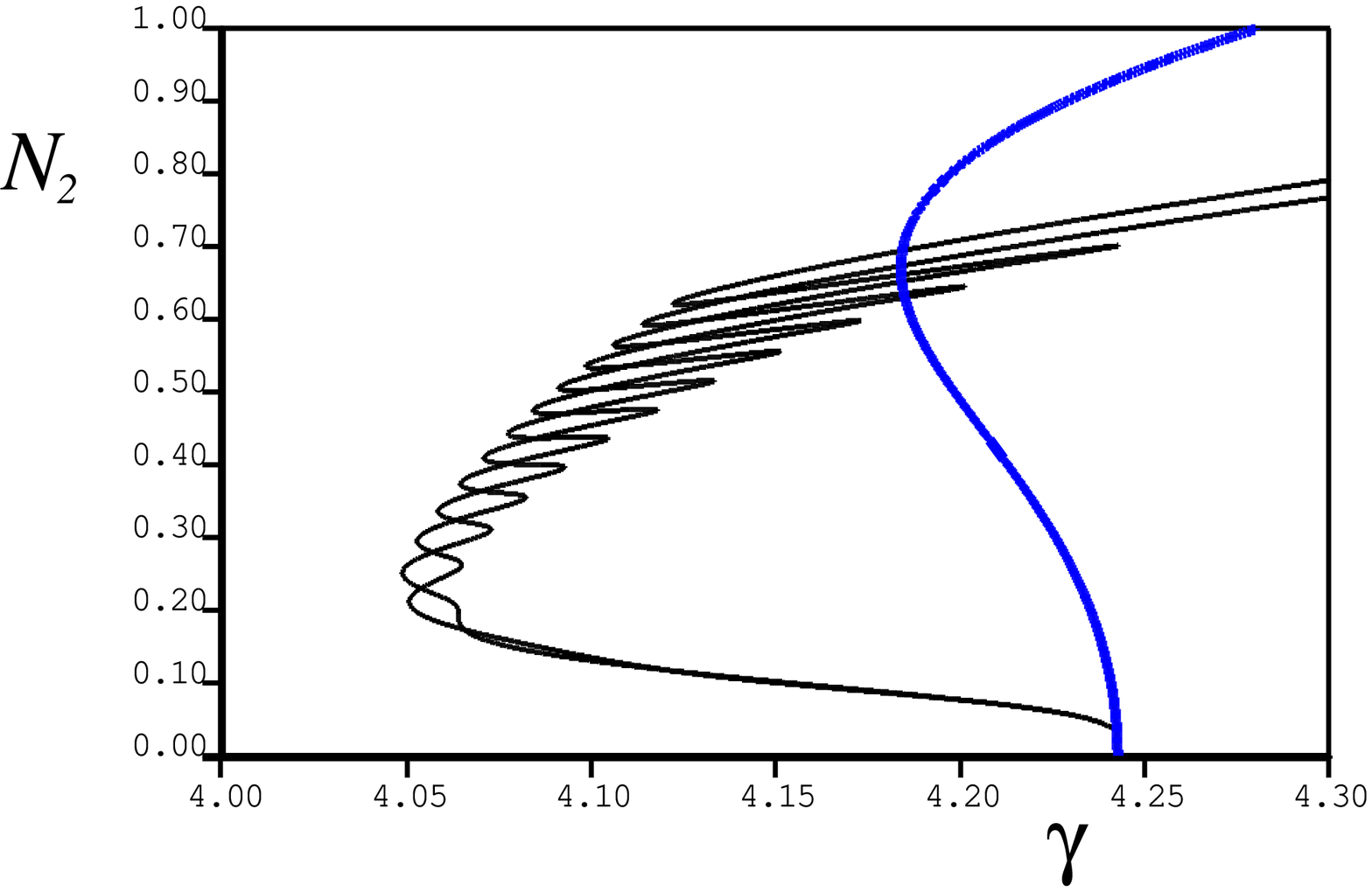}
\centerline{\phantom{a} \hspace{0.25cm}
(a) \hspace{5.5cm} (b)}
\caption{\label{fig:lsm}\small 
(a) Experimentally-determined regime diagram from Umbanhowar et al. (1996)
showing that oscillons (i.e. localized states), which exist only
in the `bubble', exist more subcritically than
extended patterns which only exist above the triangles.
$\Gamma$ is the dimensionless acceleration (driving parameter) and $f$ is
the frequency of the sinusoidal vertical motion of the granular layer.
Copyright Nature Publishing Group (1996).
(b) Bifurcation diagram for slanted snaking
in the oscillon model (Dawes \& Lilly 2009) showing localized states (intertwined
thin black curves) existing for $\gamma (\sim \Gamma)$
more negative than patterned states
(thick blue curve), corresponding to (b) around $f=25\mathrm{Hz}$. The vertical
axis $N_2$ is a solution norm.}
\ec
\end{figure}

\subsection{Large-scale modes}

In a subset of the application areas listed in
section~\ref{sec:recent}
it has been realised that the pattern-forming instability is coupled to a
large-scale mode which is neutrally stable at long wavelengths. Such
a mode arises naturally in some situations due to a conservation law
(Matthews \& Cox 2000), and cannot be ignored in a weakly
nonlinear analysis. The effect of such a neutral mode is
to `stretch out' the homoclinic snaking in parameter space,
in a way that can be captured by a more subtle asymptotic analysis
that leads naturally to Ginzburg--Landau-type equations that contain
nonlocal terms. The large-scale field causes stable 
localized states to exist over a greater parameter range than
just near the Maxwell point, and in many cases localized states
exist more subcritically than might be expected,
as illustrated in figure~\ref{fig:lsm}(b).
Recent work (Dawes 2007, 2008; Dawes \& Lilley 2009)
provides both a physical and a mathematical mechanism that
resolves differences between the traditional
homoclinic snaking bifurcation diagram~(figure~\ref{fig:snake1}a) and
the `slanted snaking' that arises in the large-scale-mode case.
Intruigingly, similar mechanisms operate in several different
application areas: models for magnetoconvection (Dawes 2007, 2008),
and nonlinear optics (Firth et al. 2007), and
experimental results obtained 
both for dielectric gas discharge (H.-G. Purwins, unpublished)
and for vertically
vibrated layers of granular material, see Umbanhowar et al. (1996) 
from which figure~\ref{fig:lsm}(a) is reproduced.
Asymptotic analysis of a model
problem (proposed by Tsimring \& Aranson 1997)
for the vertically shaken granular layer case
shows (figure~\ref{fig:lsm}b), in agreement with
experimental results (figure~\ref{fig:lsm}a),
that the localized states can exist more
subcritically than the uniform periodic pattern. The uniform periodic
pattern is
indicated
by the thick blue line in figure~\ref{fig:lsm}(b) and
bifurcates subcritically before turning around at finite
amplitude, but
the snaking curves are not constrained to lie only
between the linear instability
and the saddle-node point at the leftmost extreme
of the blue curve: they appear at lower values of the bifurcation
parameter $\gamma$ due to the large-scale mode.

\begin{figure}[!h]
\bc
\bmp{9.0cm}
\includegraphics[width=9.0cm]{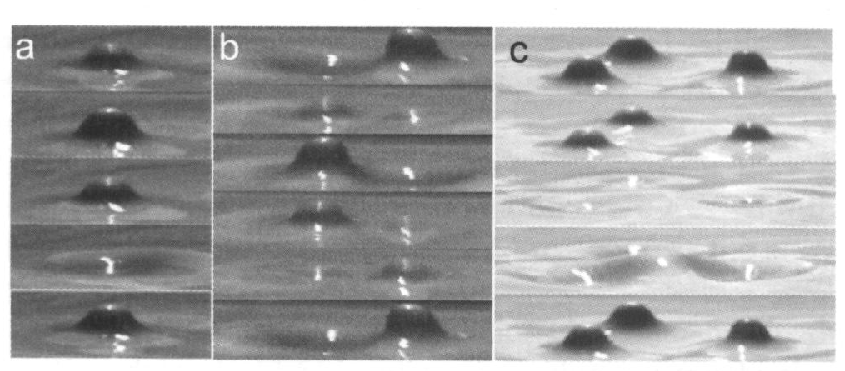}
\emp
\bmp{2.8cm}
\phantom{\tiny a}
\hspace{-0.35cm}
\includegraphics[width=2.8cm]{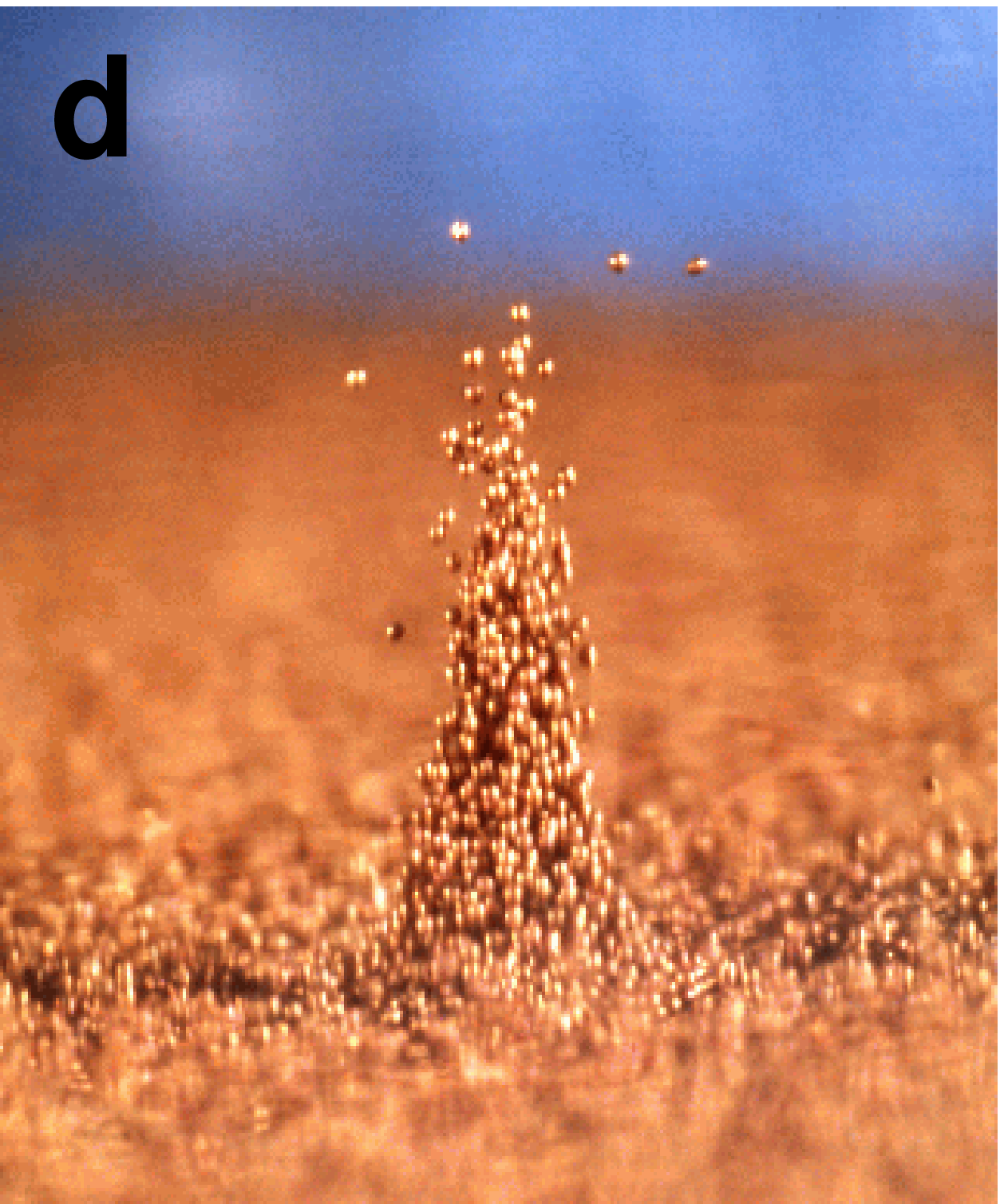}
\emp
\caption{\label{fig:ls}\small Experimentally observed oscillons
in layers of vertically shaken material: (a), (b), (c)
in a viscoelastic clay suspension
of density $\rho=1.28\mathrm{g}/\mathrm{cm}^3$; (a) shows
a single oscillon at $f=14\mathrm{Hz}$, (b) an oscillon pair at
$f=20\mathrm{Hz}$, (c) an oscillon triad at $f=25\mathrm{Hz}$.
Frames are equally spaced in time and cover two periods of the vertical
forcing, showing the subharmonic nature of the localised response
(Lioubashevski et al 1997). 
(d) Sideview of a localised state
in a layer of bronze spheres of diameter $\approx 0.16\mathrm{mm}$,
approx 17 particles deep (Umbanhowar et al 1996).
Again, the response is subharmonic: after one period of the
forcing the peak will have collapsed to a crater, re-forming a peak
after a second period of the forcing has passed.
(a) - (c) are reprinted with permission from: Lioubashevski et al (1999)
and are copyright (1999) by the American Physical Society. (d) is
copyright Nature Publishing Group (1996).}
\ec
\end{figure}

The physical mechanism for this difference is a balance between
diffusion of the large-scale quantity and nonlinear gradients
of excitation of activity. In the vertically shaken layers,
conservation of mass leads to a diffusion equation with a 
nonlinear term indicating that mass is ejected from more active
regions in the layer. This positive feedback stabilises
oscillons since this nonlinear expulsion of material
can balance diffusion even below the linear
stability threshold. In a suitable asymptotic limit this balance
can be sustained even at small amplitude, and hence a modified
weakly nonlinear analysis is possible.

\section{Applications of localized states}
\label{sec:apps}

In this section I offer three applications in which the theory
of localized states may prove extremely useful.

The general philosophical point is that
many climate change problems, illustrated here by the examples of
ocean circulations and desertification, involve
spatially extended systems which evolve smoothly until
a critical parameter value is reached where
catastrophic and irreversible change occurs: this is the idea
of a \textit{tipping point}. In the language of nonlinear dynamics, this
corresponds to the existence of a saddle-node bifurcation, 
together with hysteresis (arising from bistability) so
that after the catastrophic change there is no simple, reversible
path by which to recover the original climate state.
This is the kind of system in which
localized states exist; therefore it appears
ambitious but reasonable to suggest that properties of localized states
could form the basis of diagnostic tests for proximity to tipping
points. For example, a \textit{local} perturbation of the system
into the undesirable state produces a front between parts of the system
that are locally in the two different stable states. We then
can understand, quantitatively, how the front
will generically move in order to eliminate
the non-uniformity. Far from the tipping point
one would expect the local perturbation into the undesirable state
to vanish, while closer to the tipping point (and past the `Maxwell point'
for the system where a front between the two states would be stationary)
the undesirable state would win.
In this way, properties of the front (e.g. its velocity)
could be used to predict the distance from the current system state
to the `Maxwell point' and hence to the saddle-node
bifurcation (tipping point).

\subsection{Transition to turbulence in shear flows}
\label{sec:shear}

\begin{figure}[!h]
\bc
\includegraphics[width=6.5cm]{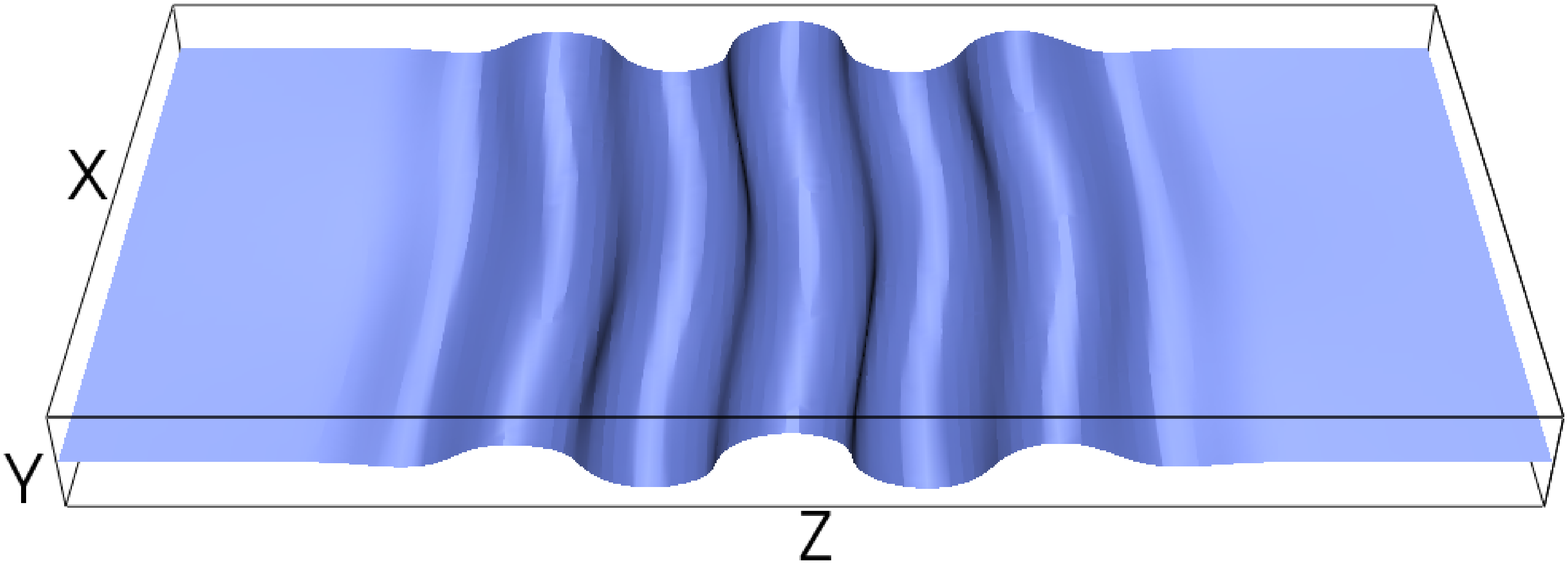}
\includegraphics[width=5.0cm]{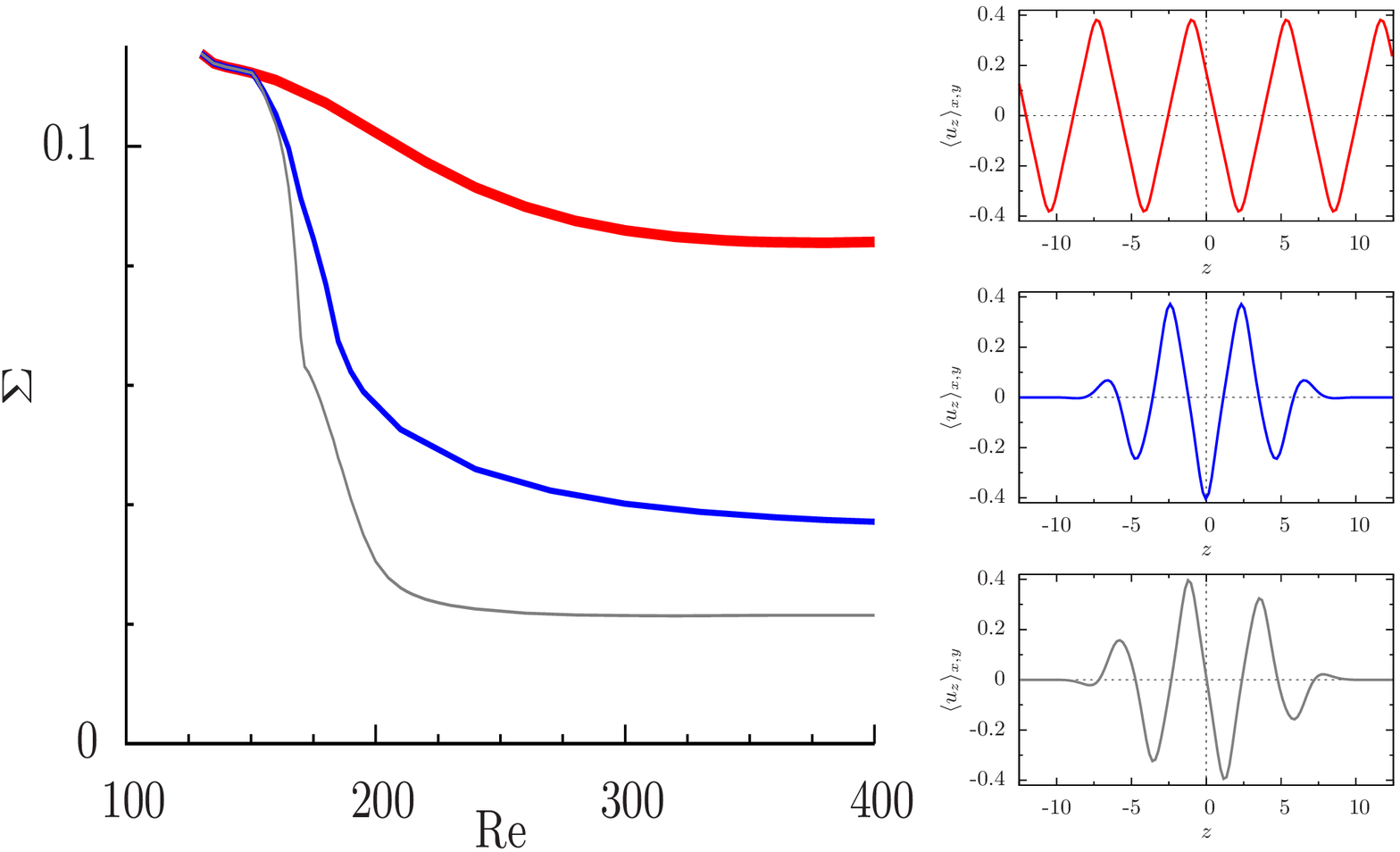}
\centerline{\phantom{a} \hspace{0.5cm} (a) \hspace{6.0cm} (b)}
\caption{\small\label{fig:shear}
(a) Surface of vanishing streamwise velocity $u$ for
an unstable steady localized state in plane Couette flow
$\mathbf{u}\equiv(u,v,w)=(y,0,0)$ between stress-free surfaces
at $y=\pm 1$, where $(u,v,w)$ are the velocity
components in, respectively, the $x$ (streamwise),
$y$ (wall-normal) and $z$ (spanwise) directions.
This very closely resembles the middle inset figure in (b)
since Couette flow is symmetric under $y \rightarrow -y$.
Reproduced with permission from
Schneider, Marinc \& Eckhardt (2009a).
(b) Typical bifurcation diagram of localized (blue, black)
and spanwise-periodic (red)
states, courtesy of Tobias Schneider.
Note the similarity of the lower two
inset figures to figure~\ref{fig:snake2}(a), (b).}
\ec

\end{figure}

One classic example of
bistability arises in the transition to turbulence from simple
shear flows, for example, plane Couette flow.
In plane Couette flow a transition from laminar flow,
to persistent turbulence
is observed experimentally and computationally
in a statistically reliable fashion with increasing Reynolds number,
without the existence of a linear
instability of the laminar state.
Exciting recent work on this problem of `transition to turbulence' has
uncovered localized states which appear to play important roles in
the structure of the collection of `edge states' -- unstable
flow structures that separate the basins of attraction
of the laminar state and the persistently turbulent one
(Schneider, Marinc \& Eckhardt 2009; Schneider, Gibson \& Burke 2010).
Understanding the details of the role of localized states
is an exciting development which will
change and vastly clarify our understanding of fluid flow at moderate
Reynolds numbers.

\subsection{The Atlantic meridional overturning circulation (MOC)}
\label{sec:moc}

The Contribution of Working Group I to the Fourth Assessment Report of the
Intergovernmental Panel on Climate Change (IPCC), Bindoff et al (2007),
discusses the importance of the MOC at length, noting the
difficulties associated with obtaining reliable data and calibrating
ocean circulation models for the MOC. The IPCC
report also affirms the influence of the MOC on global climate,
establishing its important role in climate dynamics, stating clearly
for example (chapter 5, page 397) that 
\textit{There is evidence for a link between the MOC and abrupt
changes in surface climate during the past 120 kyr...}
It is certainly clear that the MOC is a crucial determinant of European
climate since it transports heat from equator to
pole, and this is estimated to keep the
Atlantic around $4^\circ C$ warmer than it would otherwise be.

\begin{figure}[!h]
\bc
\bmp{6.5cm}
\includegraphics[width=6.5cm]{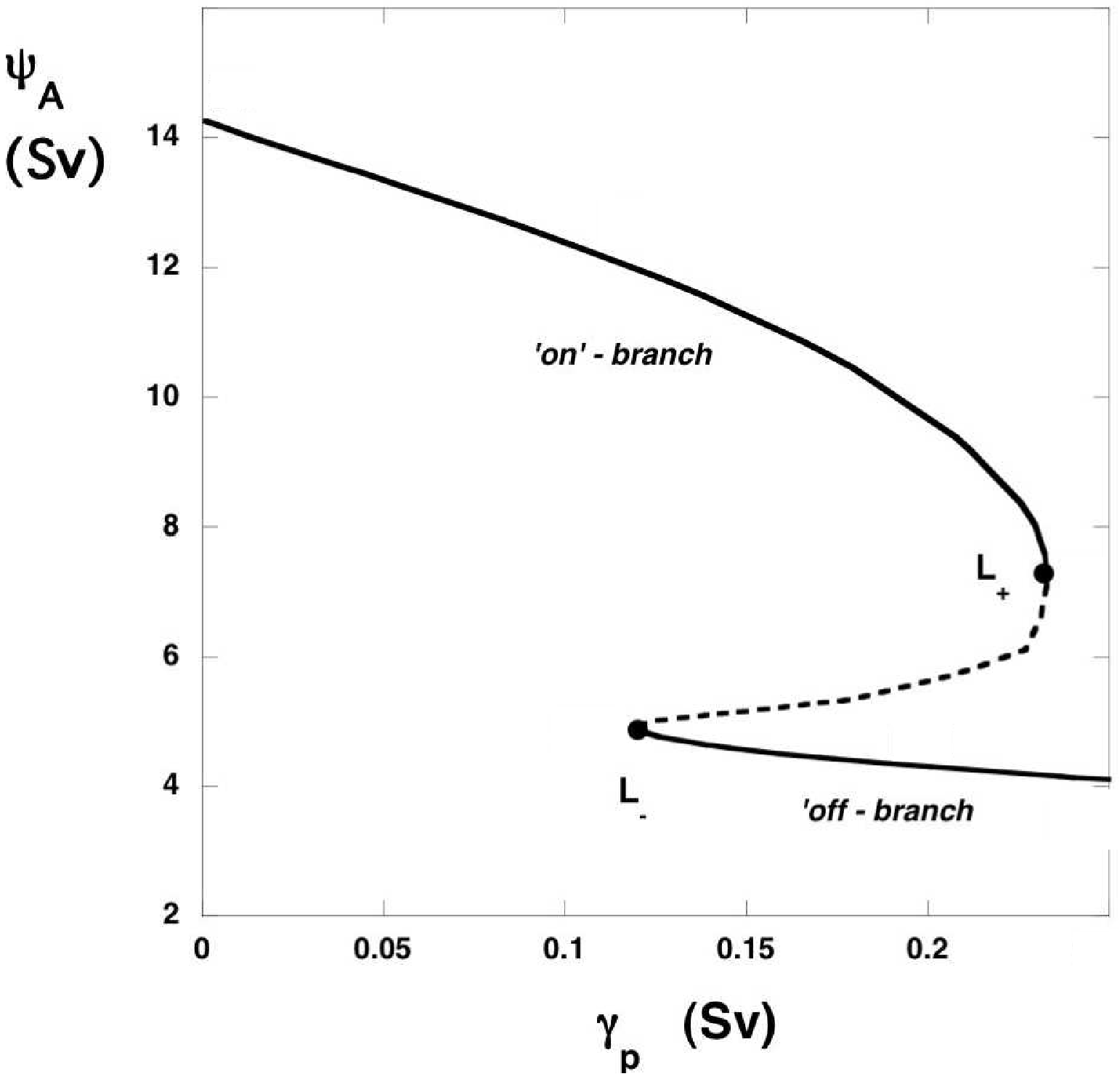}
\emp
\bmp{6.5cm}
\includegraphics[width=6.25cm]{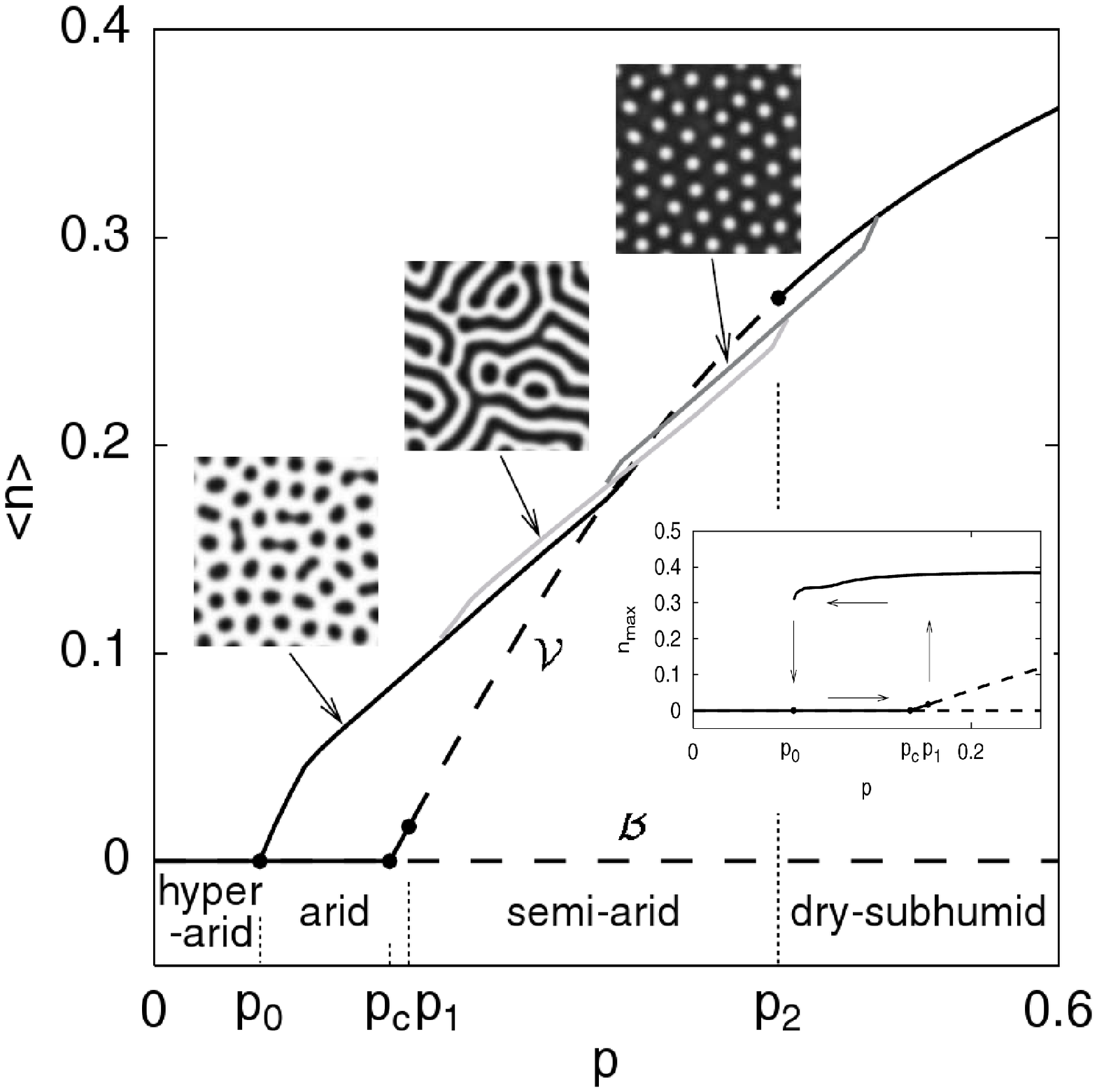}
\emp
\caption{\label{fig:moc}(a) Bifurcation diagram for the strength of the
Atlantic meridional overturning
circulation $\psi_A$ as the anomalous freshwater forcing parameter $\gamma_p$
is varied. Reproduced with permission from Huisman et al. (2009).
$Sv$ refers to the \textit{Sverdrup}, a
unit of volume flux: $1\ Sv = 10^6\,\mathrm{m}^3\mathrm{s}^{-1}$. (b)
Bifurcation diagram for the desertification
model~(\ref{eqn:d1}) - (\ref{eqn:d2})
proposed by von Hardenberg et al (2001) showing
numerically-determined ranges of
existence of different patterns, and (insert)
bistability at low precipitation rates in the range $p_0<p<p_1$.}
\ec
\end{figure}

Figure~\ref{fig:moc}(a) shows a bifurcation diagram for the
MOC computed numerically directly from a global
ocean model (Huisman et al 2009). The bistability
it shows is common to many classes of such model, originating with
Stommel (1961), and is well known (see for example
Launder \& Thompson 2008). Models suggest that
an increasing flux of freshwater from melting polar ice (i.e.
increasing $\gamma_p$) pushes the MOC towards the limit point $L_+$
which, if reached, would result in the MOC switching off abruptly, and the
European climate therefore cooling.

Simplifying the geometry of the Atlantic ocean to a
fluid-filled rectangular box, one can generate very similar
flows in the laboratory (Whitehead 1996). The equatorial region
is modelled by a heated plate attached to one side of the box, providing
a constant source of hot, and also salt-enriched, water. This hot saline
water flows across the top of the box to the opposite (`polar') side where
a stream of colder fresh water is introduced. The cold fresh water cools the
hot stream on contact, which forces it to sink since its salinity makes it
now denser than its surroundings. Hence a return flow from pole to equator
is formed at depth in the container. Throwing away the geometry and
almost all the mechanics, one can treat the problem as consisting
of flows between two separate boxes of fluid for the equatorial and polar
regions with temperature and salinity transport between them.
Let the equatorial (respectively, polar) box contain fluid at
temperature $T_e$ ($T_p$)
and salinity $S_e$ ($S_p$). `Toy model' equations for the
exchange of temperature
and salinity between the boxes were introduced by Stommel (1961):
\ba
\dot T = \eta_1 - T (1+|T-S|), & \qquad & \dot S = \eta_2 - S(\eta_3
+ |T-S|) \label{eqn:boxmodel}
\ea
where $T=T_e-T_p$ and $S=S_e-S_p$, the parameters $\eta_1$ and $\eta_2$ measure
the strengths of the thermal and freshwater forcings respectively, and $\eta_3<1$
is the ratio of relaxation times of temperature and salinity (a Lewis number).
This model robustly captures the bistability but is clearly hugely
over-simplified. Yet the bistability persists as the model is
refined, and this allows (at least in theory) the formation of fronts
and pulses. With the aid of some additional spatial inhomogeneity to
allow pinning,
localized states could be supported. At the very least, theoretical
opportunities exist to develop spatially-extended versions
of~(\ref{eqn:boxmodel}) and, for example, probe the statistics of fluctuations
that might be useful in diagnosing how close to the bistability regime, or
indeed to the point $L_+$ (see figure~\ref{fig:moc}a) the MOC currently is.

\subsection{Desertification}
\label{sec:desert}

Surprisingly similar considerations apply to models that have been proposed for
the propagation of vegetation patterns, and the reverse process, desertification
(von Hardenberg et al 2001; Meron et al 2007),
for example the pair of dimensionless PDEs
\ba
n_t & = & \frac{\gamma w}{1+\sigma w} n - n^2 - \mu n + \nabla^2 n, \label{eqn:d1} \\
w_t & = & p - (1-\rho n)w - w^2 n + \delta \nabla^2 (w-\beta n), \label{eqn:d2}
\ea
where $n(x,y,t)$ and $w(x,y,t)$ are the densities of biomass and soil water,
respectively, and $\gamma$, $\sigma$, $\mu$, $p$, $\rho$, $\delta$ and
$\beta$ are non-negative parameters. The terms on the
right-hand-side of~(\ref{eqn:d1})
describe, in order, plant growth, saturation as biomass reaches the
soil carrying capacity, mortality and predation, and reproductive
spread. The terms in~(\ref{eqn:d2}) similarly represent
precipitation, loss due to evaporation (notice that $\rho>0$ indicates
that vegetation inhibits evaporation), uptake of water by plants, and
diffusion of water through the soil, accounting for the suction of
water by plant roots.

A realistically large value for $\delta$ of around $100$ 
leads to a Turing instability in which vegetated (patterned) states
appear in addition to the trivial solution $n=0$, $w=p$ corresponding
to desert. In the low precipitation ($p_0<p<p_c$) regime, denoted
\textit{arid} on figure~\ref{fig:moc}(b), such a model
exhibits bistability
between states of arid desert (the horizontal line marked $\mathcal{B}$)
and (almost) periodic vegetation spots. The inset in figure~\ref{fig:moc}(b)
indicates the hysteresis loop: a spot pattern of vegetation cannot be
sustained when the precipitation $p$ falls below $p_0$ and the system
makes an abrupt transition to the desert state. Desert persists as $p$
is then increased up to $p_c$ at which point the desert state becomes
unstable to uniform vegetation (the line denoted $\mathcal{V}$ in the
main part of figure~\ref{fig:moc}b). $\mathcal{V}$ becomes unstable
at $p_1$ (numerical investigations indicate that this instability
is subcritical)
to patterned states: spots, then labyrithine stripes, then a uniform
vegetation pattern with holes, as $p$ increases further.

This general behaviour suggests that qualitative aspects
of the formation and dynamics of localized vegetation patches, as observed
in arid enviroments such as the Negev desert (Meron et al 2007)
may be used as diagnostic tests for proximity to sudden
desertification. The challenge is to make this a quantitative
method.

\section{Conclusions}
\label{sec:conc}

In section~\ref{sec:recent} of
this review I presented a thumbnail sketch of the current
state of our understanding of localized states, concentrating almost entirely
on the 1D case. It should be pointed out
that progress is, slowly but surely, being made in 2D and 3D as well, see
for example the papers by Lloyd et al (2008), Lloyd \& Sandstede (2009)
and Taylor \& Dawes (2009) who discuss steady localized
states, and Bode et al. (2002) who discuss localized states
that travel horizontally and interact
strongly through collisions.
In section~\ref{sec:apps} I speculated about
the usefulness of the present theory for understanding and predicting
various
abrupt changes in particular physical systems. I will finish with
brief remarks on coherent structures more generally.

Although it appears difficult to give a precise definition of a coherent
structure, the term is frequently used in describing more-or-less
fully developed turbulent flows. In such flows, large, long-lived
eddies are prominent flow features. How might the theory of localized
states be useful in this setting? Firstly, of course, one
is dealing with vortical rather than density-like coherent objects.
But nevertheless one might be able to
exploit a separation of timescales, along with a suitable
set of `effective coordinates' for the evolution of the
scale and position of these relatively large, slow-moving structures
moving over a more rapidly-evolving background flow field that can
be characterised through averaged properties. Such a reduction, at
least locally in time, might help describe the emergence and
destruction of coherent structures and thereby offer insights into
fully-developed turbulence. Elements of such an approach, with
substantial physical intuition but without
a detailed mathematical analysis,
have been employed already in nonlinear optics
(Akhmediev \& Ankiewicz 2005).
An easier place to begin pursuing
this kind of programme might be spatiotemporally complicated
flows such as `spiral defect chaos' where again the dynamics
is driven by (vertical) vorticity (Morris et al 1993).
Extensions of the 2D Swift--Hohenberg
equation to take account of vertical vorticity naturally
have a similar form to the large-scale density-like mode case
(Dawes 2008) since the 
vorticity equation is of nonlinear diffusion type, and is
neutrally stable at zero wavenumber.

\begin{acknowledgements}
The author would like to thank Helmut Brand,
John Burke, Alan Champneys, Steve Cox,
Steve Houghton, Edgar Knobloch, Gregory Kozyreff,
David Lloyd, Paul Matthews, Michael Proctor, Hans-George Purwins,
Reinhardt Richter,
Hermann Riecke, Bjorn Sandstede, Chris Taylor
and Thomas Wagenknecht for useful discussions over several
years on a wide range of topics related to this subject.
The author also gratefully acknowledges financial support
from the Royal Society through a University Research Fellowship.
\end{acknowledgements}


\end{document}